\documentstyle[12pt]{article}
\newcommand{\be}{\begin{equation}}
\newcommand{\en}{\end{equation}}
\newcommand{\bea}{\begin{eqnarray}}
\newcommand{\ena}{\end{eqnarray}}

\newcommand{\hbo}{\hbox to 1 true cm {\hfill } }
\newcommand{\tr}{\hbox{tr}}

\def\dslash{\partial\kern-.5em\slash}
\def\aslash{A\kern-.5em\slash}
\def\kslash{k\kern-.5em\slash}
\def\pslash{p\kern-.5em\slash}

\begin{document}
\vglue 1truecm

\vbox{
T95/096
\hfill August 11, 1995
}
\vbox{ hep-ph/9508307 \hfill }

\vfil
\centerline{\bf Chiral symmetry breaking in strongly coupled QED? }

\bigskip
\centerline{ K.\ Langfeld$^*$  }
\vspace{1 true cm}
\centerline{ Service de Physique Theorique, C.E.A. Saclay, }
\centerline{F--91191 Gif--sur--Yvette Cedex, France. }
\bigskip
\bigskip
\centerline{ {\bf PACS:} 11.10.GH 11.10.HI 11.15.Ex 12.20.-m }
\bigskip

\vfil
\begin{abstract}

The coupled system of renormalized Dyson-Schwinger equations for the
electron self-energy and the photon propagator are supplied with
the tree level vertex as Ansatz for the renormalized three point function.
The system
is investigated numerically. In the case of a massive electron, the
theory is ``weakly renormalizable'', i.e.\ cutoff independent for
values of the cutoff below an upper limit. In this regime of cutoff
independence, the quenched
approximation yields good results for the electron self-energy.
In the chiral limit, a logarithmic cutoff dependence of the
electron self-energy is found.
The question, whether a regime of cutoff independence with a
spontaneously broken chiral symmetry exists in strongly coupled QED,
remains open.

\end{abstract}

\vfil
\hrule width 5truecm
\vskip .2truecm
\begin{quote}
$^* $ Supported by DFG under contracts La--$932/1-1/2$. \hfill \break
\phantom{$^*$} email: {\tt langfel@amoco.saclay.cea.fr }
\end{quote}
\eject
\section{ Introduction }
\label{sec:1}

The occurrence of spontaneous chiral symmetry breaking in strongly
coupled QED is one of the most challenging issues of
non-perturbative quantum field theory nowadays. In his pioneering
work, Miranski reported a regime of cutoff independence
(CI-regime)\footnote{We do not use the notion of a ``scaling limit''
in order to avoid confusion with the term in solid state physics,
where it corresponds to the case of exact scaling (zero masses).}
corresponding to a phase of
QED with spontaneous broken chiral symmetry~\cite{mir85}. His results
are based on a study of the Dyson-Schwinger equation for the electron
self-energy, where electron loop corrections to the photon propagator
are neglected (quenched approximation). In the quenched approximation,
no renormalization of the electric charge is required, and
spontaneous symmetry breakdown occurs, if the bare charge exceeds
a critical value~\cite{mir85,bar86,co89}. Subsequently, it was
argued by Bardeen, Leung and Love that a Nambu-Jona-Lasinio type
interaction~\cite{nam61}, must be included in order to render
the renormalization of the Dyson-Schwinger equation consistent
with the known renormalization properties of the theory~\cite{bar86}.
The results of the quenched
approximation are caught into question, since vacuum polarization
effects might enforce the renormalized coupling to vanish in the
the infinite cutoff limit~\cite{landau} (triviality). First
investigations of this
effect were done by Kondo by parametrizing the effect of the vacuum
polarization~\cite{ko89}. Further insight in the issue of triviality
was gained by the important work of Rakow~\cite{rakow}. He numerically
solved the coupled set of (bare) Dyson-Schwinger equations for the
electron self-energy and the photon propagator, and therefore included
vacuum polarization effects self-consistently. He finds a second
order chiral phase transition and zero renormalized charge at the
critical point~\cite{rakow}. This seems to rule out the quenched
approximation, since the quenched approximation predicts an
interacting infinite cutoff limit.

Parallel to the studies of the truncated Dyson-Schwinger equations,
extensive lattice investigations were performed in order to
clarify the triviality problem of QED~\cite{latt94,illinois,desy}.
Lattice simulations are not bounded by any approximation, but
suffer from a small correlation length compared with the intrinsic
fermionic energy scale due to the finite lattice sizes. In order to
overcome the difficulty of a small correlation length, one fits
the lattice data to two Ans\"atze of the equation of state
(the bare electron mass as function of the fermionic condensate).
One of these Ans\"atze favors an interacting infinite cutoff
limit~\cite{illinois},
whereas the other Ansatz is compatible with triviality~\cite{desy}.
The present status is that the lattice data are not conclusive enough in
order to distinguish between the two equations of state~\cite{latt94}.

In this paper, we further develop the results of~\cite{rakow} and study the
coupled set of renormalized Dyson-Schwinger equations for the electron
self-energy and the photon propagator. The tower of Dyson-Schwinger
equations is truncated by using the tree level electron-photon-vertex
as Ansatz for the renormalized three point function. In order to introduce
our notation and to make the renormalization procedure transparent, we
first derive the renormalized Dyson-Schwinger equations (section \ref{sec:2}).
All Ans\"atze are made for renormalized quantities.
This will later turn out to be crucial in order to compare the full
results with those of the quenched approximation.
We first concentrate on the case of a massive electron (section \ref{sec:3}).
A cutoff independence is found for sufficiently small values of the cutoff.
The theory is called to be {\it weakly renormalizable}.
In contrast, the quenched approximation always predicts a CI-regime
with no upper limit on the cutoff. Although the quenched approximation
is incapable to predict the correct CI-behavior, it provides
good results for physical quantities in the CI-regime (subsection
\ref{sec:3.2}). We then focus onto the chiral limit (section \ref{sec:4}).
A logarithmic cutoff dependence is seen as well in the electron
self-energy as in the vacuum polarization.
The problem of the existence of a CI-regime characterized
by a spontaneous breakdown of chiral symmetry is still unsolved.

\section{ Renormalized Dyson-Schwinger equations }
\label{sec:2}

The generating functional for {\it bare} Green's functions of QED
in Euclidean space is given by the functional integral
\bea
W[j_\mu ^B, \eta ^B, \bar{\eta }^B ](e_B,m_B) &=& \int
{\cal D} A_\mu \; {\cal D} q \; {\cal D} \bar{q} \;
\exp \biggl\{ - \int d^4 x \; \Bigl[ L_0(x)
\label{eq:1} \\
&-& \bar{\eta }^B(x) q(x) \, - \, \bar{q}(x) \eta ^B (x) \, - \,
j_\mu ^B (x) A_\mu (x) \Bigr] \biggr\} \; ,
\nonumber
\ena
\be
L_0 (x) \;=\; \frac{1}{4e_B^2} F_{\mu \nu }[A](x) F_{\mu \nu }[A](x) \, + \,
\bar{q}(x) ( i\dslash +im_B )q(x) \, + \, \bar{q}(x) \aslash (x) q(x) \; ,
\label{eq:2}
\en
where $F_{\mu \nu }[A](x) = \partial _\mu A_\nu (x)- \partial _\nu
A_\mu (x)$ is the field strength tensor, and $e_B$, $m_B$ represent the
bare electric charge and the bare electron mass respectively.
At zero external sources, the generating functional $W[0,0,0](m^B, e^B)$
is invariant under U(1) gauge transformations of the integration
variables, i.e.
\be
q(x) \rightarrow \exp \{ i \alpha (x) \} \, q(x) \; , \hbo
A_\mu (x) \rightarrow A_\mu (x) \, + \, \partial _\mu \alpha (x) \; .
\label{eq:3}
\en
Bare Green's functions are divergent implying the theory must be
regularized and renormalized in order to make sense.
Renormalization is performed by absorbing the divergences into
the renormalization constants $Z_{2,3,e,m}$, which relate
the bare sources and the bare parameters to the renormalized ones, i.e.
\bea
\eta ^B(x) &=& Z_2^{-1/2} \, \eta (x) \; , \; \; \;
\bar{\eta }^B(x) \; = \; Z_2^{-1/2} \, \bar{\eta }(x) \; , \; \; \;
j_\mu ^B(x) \; = \; Z_3^{-1/2} \, Z_e^{-1} \, j_\mu (x) \; ,
\label{eq:4} \\
e_B &=& Z_e \, e_R \; , \; \; \; m_B \; = \; Z_m \, m_R \; .
\nonumber
\ena
The generating functional for {\it renormalized } Green's functions
is obtained from $W[j_\mu ^B, \eta ^B, \bar{\eta }^B ](e_B,m_B)$ by
replacing the bare sources and bare parameters by the
renormalized sources and the renormalized parameters respectively, i.e.
\be
W_R[j_\mu , \eta , \bar{\eta } ](e_R,m_R) \; = \;
W[Z_3^{-1/2} Z_e^{-1} j_\mu , Z_2^{-1/2} \eta , Z_2^{-1/2} \bar{\eta }]
(Z_e e_R, Z_m m_R) \; .
\label{eq:5}
\en
The renormalized Green's functions are obtained by taking functional
derivatives of $W_R$ (\ref{eq:5}) with respect to the corresponding
external sources. For example, the renormalized electron propagator
$S_R$ and the renormalized photon propagator $D_{\mu \nu }^R$ are
given by
\be
S_R(x,y) \; = \; \frac{ \delta \ln W_R[j_\mu , \eta , \bar{\eta } ] }{
\delta \eta (x) \, \delta \bar{\eta }(y) }
\vert _{j, \eta , \bar{\eta } = 0 } \; , \; \; \;
D_{\mu \nu }^R (x,y) \; = \; \frac{ \delta \ln W_R[j_\mu , \eta ,
\bar{\eta } ] }{ \delta j_\mu (x) \, \delta j_\nu (y) }
\vert _{j, \eta , \bar{\eta } = 0 } \; .
\label{eq:5a}
\en
After a change of the integration variables, the generating functional
(\ref{eq:5}) can be cast into
\bea
W_R[j_\mu , \eta , \bar{\eta } ](e_R,m_R) &=& \int
{\cal D} A_\mu \; {\cal D} q \; {\cal D} \bar{q} \;
\exp \biggl\{ - \int d^4 x \; \Bigl[ L(x)
\label{eq:6}  \\
&-& \bar{\eta }(x) q(x) \, - \, \bar{q}(x) \eta (x) \, - \,
j_\mu (x) A_\mu (x) \Bigr] \biggr\} \; ,
\nonumber
\ena
\bea
L (x) &=& \frac{Z_3}{4e_R^2} \, F_{\mu \nu }[A](x) F_{\mu \nu }[A](x)
\, + \, Z_2 \, \bar{q}(x) i\dslash q(x) \, + \, Z_0 \, im_R \bar{q}(x) q(x)
\label{eq:7} \\
& + & Z_1 \, \bar{q}(x) \aslash (x) q(x) \; ,
\ena
where we have introduced
\be
Z_1 \; = \; Z_2 \, Z_3^{1/2} \, Z_e \; , \hbo
Z_0 \; = \; Z_2 \, Z_m \; .
\label{eq:8}
\en
The generating functional $\Gamma $ for renormalized
one-particle irreducible functions (in the following vertex functions)
is obtained from $W_R$ (\ref{eq:5}) by a Legendre transformation, i.e.
\be
\Gamma [{\cal A}_\mu , \psi , \bar{\psi }] \; = \;
\ln W_{R}[j_\mu , \eta, \bar{\eta }] \; - \; \int d^4x \; \left[
j_\mu (x) {\cal A}_\mu (x) + \bar{\eta }(x) \psi (x) +
\bar{\psi }(x) \eta (x) \right] \; ,
\label{eq:9}
\en
where the external sources are implicitly related to the fields
${\cal A}$, $\psi $, $\bar{\psi }$ by
\be
{\cal A}_\mu (x) = \frac{ \delta \ln W_R [j_\mu ,\eta , \bar{\eta } ] }{
\delta j_\mu (x) } \; , \; \;
\psi (x) = \frac{ \delta \ln W_R [j_\mu ,\eta , \bar{\eta } ] }{
\delta \bar{\eta } (x) } \; , \; \;
\bar{\psi } (x) = \frac{ \delta \ln W_R [j_\mu ,\eta , \bar{\eta } ] }{
\delta \eta (x) } \; .
\label{eq:10}
\en
Below we will use the renormalized electron photon vertex, which is
defined by
\be
\Lambda _\mu (x,y,z) \; = \; \frac{ \delta \Gamma [{\cal A}_\mu , \psi ,
\bar{\psi }] }{ \delta {\cal A}_\mu (x) \, \delta \psi (y) \,
\delta \bar{\psi }(z) } \vert _{{\cal A}_\mu , \psi , \bar{\psi } = 0} \; .
\label{eq:11}
\en
Exploiting the fact that the functional integral $W_R [j_\mu ,\eta ,
\bar{\eta } ]$ (\ref{eq:6}) is not changed by a shift of the integration
variables $A_\mu $, $q$, $\bar{q} $ generates the {\it renormalized }
Dyson-Schwinger equations~\cite{itz80}. In particular, one finds
\bea
\left( S_R(x,y) \right) ^{-1} &=& \left[ Z_2 \, i \dslash
+ Z_0 \, im_R \right] \, \delta (x-y)
\label{eq:12} \\
&-& Z_1 \int d^4z \; d^4w \; \gamma _\mu D_{\mu \nu }^R (x,z) S_R(x,w)
\Lambda _\nu (z,w,y) \; ,
\nonumber \\
\delta _{\mu \nu } \delta (x-y) &=& \frac{Z_3}{e_R^2}
\left( - \partial ^2 \delta _{\mu \alpha } + \partial _\mu
\partial _\alpha  \right) D_{\alpha \nu }^R (x,y)
\label{eq:13} \\
&+& Z_1 \int d^4z \; d^4w \; d^4v \; \tr \left\{ \gamma _\mu S_R(x,z)
\Lambda _\alpha (v,z,w) S_R(w,x) \right\} D_{\alpha \nu }^R (v,y) \, .
\nonumber
\ena
Throughout this paper, we will truncate the tower of Dyson-Schwinger
equations by making an Ansatz for the renormalized vertex function
$\Lambda _\nu $. In this case, the renormalized electron propagator and
the renormalized photon propagator can be obtained by solving
the coupled system (\ref{eq:12},\ref{eq:13}).

In order to investigate the consistency of the Ansatz, one studies
the Ward-Takahashi identities~\cite{itz80}, which provide
relations among Green's functions induced by gauge invariance.
One first realizes that $Z_1=Z_2$ must hold in the renormalized
Lagrangian (\ref{eq:7}) in order to preserve the invariance under the
transformation (\ref{eq:3}).
Combining (\ref{eq:8}) with (\ref{eq:4}), one finds that
the electric charge is renormalized by the photon wave function
renormalization constant, i.e. $e_R= Z_3 ^{1/2} e_B$. If the bare
charge $e_B$ acquires a finite value in the limit of large cutoff
(as suggested by the quenched approximation), and if
further this limit enforces $Z_3$ to vanish, then the theory
is only consistent with a zero renormalized electric charge. This scenario
is referred to as triviality of QED in the literature.

Exploring the fact that the generating
functional $W_R$ (\ref{eq:6}) is invariant under small gauge
rotations (\ref{eq:3}), one obtains
\be
\partial _\mu ^{(x)} \, \Lambda _\mu (x,y,z) \; = \;
i S_R^{-1}(y,x) \, \delta (x-z) \; - \; i S_R^{-1}(x,z) \, \delta (x-y) \; .
\label{eq:14}
\en
Once the coupled system (\ref{eq:12},\ref{eq:13}) was solved for
a particular choice of the renormalized vertex $\Lambda _\mu $,
one inserts the solution for $S_R$ into (\ref{eq:14}) in order to
check the accuracy of the Ansatz for the vertex function $\Lambda _\mu $.

\medskip

In the following, we will work in Landau gauge and set $Z_1=Z_2$ as
imposed by gauge invariance. We will study the Ansatz
\be
\Lambda _\mu (z,x,y) \; = \; \gamma _\mu \; \delta (z-x) \;
\delta (z-y) \;
\label{eq:15}
\en
for the renormalized vertex function, which is the tree level
electron-photon vertex. Recently, the full one-loop QED vertex was
obtained~\cite{kiz95}. The results might provide the structure
for a more general ansatz than (\ref{eq:15}).
Note that the choice of the renormalized
vertex is the only freedom we have. There is no further possibility to
argue in favor of a four fermion interaction first introduced in~\cite{bar86}.
Note also that a four fermion interaction arises and can be naturally
incorporated in renormalized Dyson-Schwinger equations in the dual
formulation of QED~\cite{sugamoto}.

The coupled
Dyson-Schwinger equations are then solved by parametrizing the
renormalized electron propagator and the renormalized photon
propagator in momentum space by
\be
\widetilde{S}_R(p) \; = \; \frac{1}{ F(p^2) \pslash \, + \, i
\Sigma (p^2) } \; , \hbo
\widetilde{D}^{R}_{\mu \nu }(p) \; = \; \frac{ 4 \pi D_R(p^2) }{ p^2 } \;
\left( \delta _{\mu \nu } \, - \, \frac{ p_\mu p_\nu }{p^2} \right) \, .
\label{eq:16}
\en
Up to the occurrence of the renormalization constants $Z_{1,3}$,
the derivation yields the same results as reported in~\cite{rakow}.
The first Dyson-Schwinger equation is straightforwardly reduced to
two equations to determine $F(p^2)$ and $\Sigma (p^2) $. Using a
sharp O(4)-invariant cutoff $\Lambda $ to regularize the momentum
integration, these equations are
\bea
\Sigma (p^2) &=& m_0 \, + \, Z_1 \frac{3}{ 2 \pi ^2 } \int _0^{\Lambda ^2 }
dk^2 \; k^2 \; \int _0^\pi d\theta \; \sin ^2 \theta \;
\frac{ \Sigma (k^2) }{ s(k^2) } \, \frac{ D_R(q^2) }{ q^2 } \; ,
\label{eq:17} \\
F(p^2) &=& Z_1 \, \Biggl[ 1 \; + \; \frac{1}{\pi ^2 p }
\label{eq:18} \\
&& \times \int _0^{\Lambda } dk \; k^4 \;
\int _0^\pi d\theta \; \sin ^2 \theta \; \frac{ F (k^2) }{ s(k^2) }
\frac{ D_R(q^2) }{q^4} \left\{ 3 q^2 \cos \theta - 2 k p \sin ^2 \theta
\right\} \, \Biggr] \; ,
\nonumber
\ena
where we have introduced $m_0 = Z_m m_R $, $s(k^2) =
F(k^2) k^2 + \Sigma (k^2) $ and $q^2 = k^2 + p^2 -2 k p \cos \theta $.
The equation for the photon propagator (\ref{eq:13}) needs more thoughts.
The vacuum polarization $\Pi _{\mu \nu }(p)$ due to the electron loop
is transverse, i.e. $p_\mu \, \Pi _{\mu \nu }(p) =0$, if the
regularization prescription does not violate gauge invariance.
Unfortunately, the sharp O(4)-invariant cutoff prescription spoils gauge
invariance implying that the vacuum polarization $\Pi _{\mu \nu }(p)$
acquires spurious terms proportional to $\delta _{\mu \nu }$, which
are, in addition, quadratic divergent. On the other hand, regularization
schemes consistent with gauge invariance are very time consuming when
solving the coupled equations (\ref{eq:12},\ref{eq:13}) numerically.
In order to circumvent this problem, one starts calculating
$\Pi _{\mu \nu }(p)$ using a regularization scheme which respects gauge
invariance (e.g.~Schwinger proper time, Pauli-Villars regularization).
The result for $\Pi _{\mu \nu }(p)$ is the transverse projector times a
function depending on $p^2$. In this function, one first introduces a second
regularization which corresponds to the O(4)-invariant cutoff scheme, and
then removes the regulator of the scheme which is compatible with gauge
invariance. The final result allows a fast numerical treatment
and is consistent with the transversality of the vacuum polarization
$\Pi _{\mu \nu }(p) $. In the limit of a large cutoff,
all regularization schemes
which respect the symmetries of the model are supposed to yield the
same result for physical quantities.
In agreement with the result stated in~\cite{rakow},
this lengthy procedure yields
\bea
\frac{1}{D_R(p^2)} &=& \frac{Z_3}{\alpha _R} \; - \; Z_1 \frac{2}{3 \pi ^2}
\label{eq:19} \\
&& \times \int _0 ^{\Lambda ^2 } dk^2 \; k^2 \, \int _0^\pi d\theta \;
\sin ^2 \theta \; \frac{ F(q_+^2) \, F(q_-^2) }{
s(q_+^2) \, s(q_-^2) } \, \left\{ \frac{ k^2}{p^2} ( 8 \cos ^2 \theta
-2 ) - 3/2 \right\} \; ,
\nonumber
\ena
where $\alpha _R = e_R^2/4\pi $ and $q_{\pm} = (k \pm p/2)^2$.

\section{ QED with massive electrons }
\label{sec:3}

In order to calculate the renormalization constants $Z_{1,3}$ and $m_0$
as function of the cutoff, we impose renormalization conditions. In order
to fix $Z_1$, we demand that the inverse renormalized electron propagator
has the canonical kinetic term, i.e. $F(0)=1$. Our conventions imply
that the residue of the renormalized photon propagator $D_R(0)$ can be
identified with $\alpha _R$. Providing a value for $\alpha _R$
determines the constant $Z_3$. Finally, we set the scale of the
electron self energy $\Sigma (p^2)$ by the constraint $\Sigma (0)
=\Sigma _0$, which yields a condition to calculate $m_0$.
Although $m_0$ might tend to zero for infinite cutoff, the renormalized
(current) mass $m_R$ is non-zero. The behavior of $m_0$ in the quenched
approximation might serve as an illustrative example~\cite{bar86,co89}, i.e.
\be
m_0 \; = \; \left( \frac{\mu }{\Lambda } \right)^w \; m_R(\mu) \; ,
\hbox to 2 true cm {\hfil with \hfil }
w=1-\sqrt{1- \frac{3e^2}{4\pi ^2 } } \; ,
\en
where $\mu $ is an arbitrary renormalization point.
Since $m_0$ will turn out to be different from zero for any fixed value
of the cutoff, the renormalized (current) mass $m_R$ is also non-zero
implying that the above set of renormalization conditions corresponds to
the case of a massive electron. The chiral symmetry is explicitly broken.

\subsection{ Numerical results }
\label{sec:3.1}

The set of coupled equations (\ref{eq:17},\ref{eq:18},\ref{eq:19})
for the functions $F(p^2)$, $\Sigma (p^2)$, $D_R(p^2)$ was
numerically solved by iteration. The integrals were done by
Simpson's integration with a step-size control. A CI-regime
is found for sufficiently small cutoff. A representative result
is shown in figure~1 for $\alpha _R = D_R(0) = 0.35$. The function
$F(p^2)$ is weakly $p$-dependent. The maximum deviation from one
occurs for $p^2=\Lambda ^2$ and is $0.04776$ for $\ln \Lambda ^2
/ \Sigma _0^2 =16$. The self-energy
roughly stays constant for $\ln p^2/ \Sigma _0^2 \stackrel{<}{\sim } 4 $
and smoothly decays afterwards. The vacuum polarization also stays
constant up to approximately the same momentum and then decays
according to
\be
\frac{1}{D_R(p^2)} \; = \; c_1 (\alpha _R) \; - \;
c_2(\alpha _R) \, \ln \Lambda ^2/ p^2 \; ,
\hbox to 2cm {\hfil for \hfil }
(\ln p^2 / \Sigma _0^2 > 4 ) \; ,
\label{eq:20}
\en
where $c_1(0.35) \approx 2.97 $ and where $c_2 \approx 1/\pi ^2$ for
a wide range of renormalized couplings. The renormalization constants
are given in the table below.

\begin{center}
\begin{tabular}{|c|c|c|c|} \hline
$\ln \Lambda ^2 / \Sigma _0^2 $ & $Z_1$ & $Z_3$ & $m_0/\Sigma _0$ \\ \hline
   7  & 1.0097  &  0.75537  &  0.49437 \\
  10  & 1.0183  &  0.64295  &  0.33425 \\
  13  & 1.0305  &  0.52927  &  0.20643 \\
  16  & 1.0497  &  0.41307  &  0.10936 \\ \hline
\end{tabular}
\end{center}

The renormalization group flow is shown in figure~\ref{fig:2} for
$\alpha _R = 0.35 $. The vertex renormalization constant $Z_1$
(which is identical to the wave function renormalization constant $Z_2$
of the electron) stays close to one. The photon wave function
renormalization constant almost decays logarithmically.
The behavior of the bare mass $m_0$ corresponds to a powerlaw
decay with logarithmic corrections.

In order to check the accuracy
of the Ansatz (\ref{eq:15}) for the vertex, we study the Ward
identity (\ref{eq:14}) in momentum space and at zero photon momentum, i.e.
\be
\widetilde{\Lambda }_{\mu }(0,p) \; = \; \frac{ \partial }{
\partial _\mu } S^{-1}(p) \; .
\label{eq:20a}
\en
Using the parametrization (\ref{eq:16}) of the electron propagator and
neglecting derivatives of the functions $F(p^2)$ and $\Sigma (p^2)$, the
Ward identity (\ref{eq:20a}) becomes
\be
\widetilde{\Lambda }_{\mu }(0,p) \; = \; F(p^2) \, \gamma _\mu \; .
\label{eq:20b}
\en
The deviation of $F(p^2)$ from unity measures the violation
of the ward identity due to the Ansatz (\ref{eq:15}) for the vertex.
The numerical result shows that this violation is small, since
the deviation of $F(p^2)$ from unity is always beyond 5\% for
the values of the cutoff used in table 1.

Increasing the cutoff for fixed renormalized coupling $\alpha _R$, one
observes a critical upper limit $\Lambda _C$ of the cutoff.
Beyond this critical
value, no solution of the set of equations
(\ref{eq:16},\ref{eq:17},\ref{eq:18}) was found. This result can be
anticipated from the asymptotic behavior of the vacuum polarization
(\ref{eq:20}). For a fixed (positive) value of the renormalized
coupling $\alpha _R =D_R(0)$, equation (\ref{eq:20}) cannot be
satisfied for an arbitrarily large cutoff $\Lambda $.
Figure~3 shows the relation between the maximal possible
renormalized coupling and the critical cutoff.
The results indicate that in the limit $\Lambda \rightarrow \infty $
a solution of the coupled
Dyson-Schwinger equations only exists for $\alpha _R \rightarrow 0$.
This phenomenon is referred to as triviality of massive QED.
Note, however, that for values of the cutoff smaller than the critical
value the electron self-energy is cutoff independent. The theory is called
weakly renormalizable. For moderate values of the renormalized
coupling, the theory allows for a large value of the critical cutoff
and therefore for a cutoff many times bigger than the typical
energy scale set by the mass of the electron. This implies that
weakly renormalizable QED is perfectly compatible with the observations
in nature. The lack of a cutoff independence at high momentum does no harm,
since QED
in the real world is embedded in the Weinberg-Salam model at high energies.

\subsection{ The quenched approximation }
\label{sec:3.2}

Neglecting fermion loop effects in the photon propagator is called
quenched approximation. In our formulation, this corresponds to
replacing (\ref{eq:19}) by
\be
\frac{1}{D_R(p^2)} \; = \; \frac{Z_3}{\alpha _R} \; .
\label{eq:21}
\en
The vacuum polarization $D_R(p^2)$ is momentum independent, and from
the renormalization conditions (see beginning of section \ref{sec:3})
one immediately obtains $Z_3=1$. For a constant vacuum polarization
$D_R$, the angle integral in (\ref{eq:18}) can be done explicitly and
yields zero. This implies that $F(p^2)=Z_1$ is also constant, and
taking into account the renormalization conditions, we have $Z_1=1$.
The only remaining non-trivial equation is the integral equation for
the electron self-energy (\ref{eq:17}). For standard QED (massless photon),
this integral equation can be transformed into a non-linear
differential equation with appropriate boundary
conditions~\cite{mir85,bar86,ko89,co89}.
In the table below, we compare the
renormalization constants of the full and the quenched approach
for $\alpha _R=0.35$ and a cutoff in the CI-region
$\ln \Lambda ^2 /\Sigma _0^2 = 16$.

\begin{center}
\begin{tabular}{|c|c|c|c|} \hline
         & $Z_1$ & $Z_3$ & $m_0/\Sigma _0$ \\ \hline
quenched & 1 & 1 & 0.23172 \\
full     & 1.0497  &  0.41307  &  0.10936 \\ \hline
\end{tabular}
\end{center}

Large deviations are found. Note, however, that renormalization constants
are not physical observables, and the quenched approximation might
improve when physical quantities are studied. Figure~\ref{fig:4}
shows the electron self-energy in the quenched approximation in
comparison with the full result.
One finds that the quenched approximation yields good results for the
electron self-energy at least for small momentum $p$. This behavior
can be understood by a scaling argument. Due to cutoff independence,
the momentum
dependence of the electron-self-energy for a small momentum $p$
and for cutoff slightly below the upper critical value is also obtained
in a calculation with a cutoff far beyond the critical cutoff and
momentum $p$ comparable to this cutoff (see figure 1).
At small cutoff, however, the bare electron mass $m_0$ is large
(see table 1), and
the electron loop contributing to the inverse vacuum polarization
in (\ref{eq:19}) is negligible, hence the quenched approximation
is good.

If the cutoff exceeds the upper critical limit, a solution
of the full equations ceases to exist, whereas the quenched
approximation still predicts a solution. This implies that
polarization effects are important to address the CI-behavior
of the model.

We conclude that within the quenched approximation one cannot decide
whether a CI-regime exists or not. If a CI-regime exists,
the quenched approximation may be a good approximation for
physical observables, although non-observables (e.g.\ renormalization
constants) might turn out to be completely different.

\section{ CI-violation in chiral symmetric QED }
\label{sec:4}

In the following, we will study the chiral limit $m_R=0$ (implying
$m_0 \equiv 0$) of QED, which requires a new set of renormalization
conditions. Again we demand that the residue of the electron
propagator is one, i.e.\ $F(0)=1$, which fixes $Z_1$.
We are interested in a phase with spontaneously broken
chiral symmetry. We therefore insist on $\Sigma (0)= \Sigma _0$,
which must now be accomplished by a choice of $Z_3$ since $m_0$
is now identical zero. All renormalization constants are fixed.
The renormalized coupling $\alpha _R = D_R(0)$ will be self-consistently
calculated.

We numerically solved the coupled equations
(\ref{eq:17},\ref{eq:18},\ref{eq:19}) for this set of renormalization
conditions. The result for the electron self-energy $\Sigma (p^2)$
and the vacuum polarization $D_R(p^2)$ is shown in figure~5,
which should be compared with figure 1. We find a logarithmic
dependence on the cutoff. Although the qualitative behavior of
$1/D_R(p^2)$ is qualitatively the same as in the case of massive
electron, the plateau at small $p^2$ does not stabilize, but
continuously increases with increasing cutoff. One might think that
this CI-violation in the self-energy is due to the change of the
plateau of $D_R(p^2)$ when the cutoff is varied.
One might therefore be tempted to parametrize
the vacuum polarization by
\be
\frac{1}{D_R(p^2)} \; = \; c_3 (\alpha _R) \; - \;
c_4(\alpha _R) \, \ln \Lambda ^2/ p^2 \; ,
\hbox to 2cm {\hfil for \hfil }
\Lambda ^2 \rightarrow \infty
\label{eq:22}
\en
and to search for a CI-regime by solving the reduced set of
equations (\ref{eq:17},\ref{eq:18}). However, it turns out that
the CI-violation in the electron self-energy qualitatively
remains the same.

Two possible explanations of this cutoff dependence are immediate:
the theory is trivial in a sense that the limit $\Lambda \rightarrow
\infty $ is only compatible with $\alpha _R=0$.
In this case, the interaction is not strong enough to spontaneously
break chiral symmetry, and one cannot keep the renormalization
condition $\Sigma (0)=\Sigma _0 \not= 0$.
A second possible explanation is that in the chiral limit
the solutions are sensitive to the infrared behavior of the integrals
in (\ref{eq:17},\ref{eq:18},\ref{eq:19}). The momentum independent
Ansatz for the renormalized
vertex (\ref{eq:15}) might be to crude for the chiral limit and induces
the logarithmic CI-violation. An improved Ansatz for the vertex
function might provide a CI-regime with a spontaneous broken
symmetry. None of these two explanations can be favored without
further studies beyond the Ansatz (\ref{eq:15}) for the vertex
studied here.

\section{Conclusions}
\label{sec:5}

We numerically studied the coupled set of renormalized
Dyson-Schwinger equations for the electron and the photon propagator
using a tree level electron-photon-vertex as an Ansatz for the
renormalized vertex function. We imposed the constraint $Z_2=Z_1$
between the electron wave function renormalization constant $Z_2$
and the vertex renormalization constant $Z_1$ as required by gauge
invariance. The violation of the Ward identity is smaller than 5\%.

In the case of a massive electron (explicitly broken chiral symmetry)
and fixed renormalized coupling $\alpha _R$, a regime of cutoff
independence (CI-regime) was found, if the cutoff is below an upper
critical value. The theory is weakly renormalizable.
For larger values of the cutoff than the critical value,
no solution of the Dyson-Schwinger equations exists.
The results indicate that the infinite cutoff limit is only compatible with
a zero renormalized charge. In the CI-regime, the quenched
approximation yields reasonable results for the electron self-energy,
although the values of the renormalization constants completely
differ from the full result.

In the case of chiral symmetry (zero renormalized mass of the electron)
a logarithmic dependence on the cutoff was found. The question,
whether this cutoff dependence is due to the crude Ansatz for
the renormalized vertex or induced by triviality, cannot be answered
at the present stage of investigations. The question, whether
a CI-regime with spontaneously broken chiral symmetry exists in
strongly coupled QED, cannot be answered from the viewpoint of
the truncated Dyson-Schwinger equations supplied with tree level
vertex.

\bigskip
{\bf Acknowledgements: }
I thank J.-B.~Zuber for helpful discussions.

\newpage
\centerline{ \bf \large Figure captions }
\vspace{2 true cm }

\vspace{.5 cm}

   Figure 1: CI-behavior: the electron self-energy $\Sigma (p^2)$ and the
   vacuum polarization $D_R$ as function of the
   momentum $p^2$   for several values of the cutoff $\Lambda $ in units
   of $\Sigma _0 = \Sigma (0)$.

\setcounter{figure}{1}
\begin{figure}[h]
\vspace{.5cm}
\caption{ The renormalization group flow for $\alpha _R = 0.35$;
   $Z_1(\Lambda )$ short dashed line, $Z_3(\Lambda )$ long dashed line,
   $m_0(\Lambda )/\Sigma _0 $ solid line. }
\label{fig:2}
\end{figure}

\vspace{.5 cm}

   Figure 3: The maximal possible renormalized coupling $\alpha _R$
   as function of the critical cutoff $\Lambda _c$ (left).
   The ratio $Z_3/\alpha _R$ at the critical cutoff (right).

\vspace{.5 true cm }

\setcounter{figure}{3}
\begin{figure}[h]
\vspace{.5cm}
\caption{ The electron self-energy $\Sigma (p^2)$ in the quenched
   approximation compared with the full result for $\alpha _R=0.35$
   and $\Lambda ^2 = 8.9 \, 10^{6} \, \Sigma _0 $. }
\label{fig:4}
\end{figure}

\vspace{.5 cm}

   Figure 5: CI-violations:
   the electron self-energy $\Sigma (p^2)$ and the
   vacuum polarization $D_R$ as function of the
   momentum $p^2$   for several values of the cutoff $\Lambda $ in units
   of $\Sigma _0 = \Sigma (0)$ in the chiral limit $m_R=0$.

\end{document}